Accuracy of Refractive Index Spectroscopy by Broadband Interferometry

Yago Arosa, Carlos Damián Rodríguez-Fernández, Alejandro Doval, Elena López Lago and Raúl de la Fuente

Nanomateriais, Fotónica e Materia Branda, Departamento de Física Aplicada, Universidade de Santiago de Compostela, Spain.

Abstract

Uncertainties of refractive and group index in dispersion measurement by spectrally resolved white light interferometry are deeply analyzed. First, the contribution to uncertainty of the different parameters affecting both indices is identified. Afterwards, results are presented for a 1.5 mm thick fused silica sample over a broad spectral range, from 400 to 1000 nm, and the effects that mostly deteriorate the measurement accuracy are established. Finally, the different contributions are quadratically combined to determine the total uncertainty of the two indices.

1. Introduction

The possibility of using spectrally resolved white light interferometry (SRWLI) as a refractometric technique was already suggested in the early 1990s [1]. The first studies [2, 3], focused on the measurement of the refractive index and its dispersion curve, $n(\lambda)$, in a relatively wide spectral range (tens of nanometers). Soon after, these works were extended to incorporate the measurement of other dispersion parameters, mainly chromatic dispersion, $dn/d\lambda$, and group index, $d(\sigma n)/d\sigma$, where $\sigma = 1/\lambda$ is the wavenumber, although some authors evaluated higher order dispersion parameters [4, 5]. All these studies aim to measure the refractive index or some kind of spectral derivative by combining spectroscopy with interferometry, and we will refer to these techniques as Refractive Index Spectroscopy by Broadband Interferometry (RISBI). Like other techniques based on SRWLI, to apply RISBI we basically need three elements: a broadband source, an interferometer, and a spectrometer. RISBI was used to measure the dispersion of reference materials, both isotropic [6-10] and anisotropic [11-13], to measure dispersion in fibers [14-17], to determine the group delay of dispersive mirrors [18, 19], to measure ocular dispersion [20], to calculate thermo-optical coefficients [21], visualization of thermal lens effect [22] or to model the dispersion of families of fluids [23, 24]. Furthermore, while the first works applied RISBI in the visible range, dispersion results in the near infrared [25, 26] or even the UV range [27] can be found in the literature.

Of course, the validity of the obtained results was analyzed in many works, but the analysis was often based on comparison with the values obtained using other methods. However, in some works, [1, 26, 28], the precision or the uncertainty of the measurements was alluded to, but always in a simplified way, and only regarding to refractive index. In this work, we aim to evaluate the accuracy of the measurement of the refractive and group index, and its dispersion in a broad and rigorous way. To perform the study, we rely on measurements taken with our experimental RISBI system. However, the analysis can be easily extrapolated to any other RISBI device. In the next section, the different parameters that affect the measurements are identified and the fundamental formulas to calculate the two indices and their corresponding uncertainties are presented. In Section 3, the uncertainties associated to each of these parameters are modeled and evaluated separately, and, in Section 4, the results obtained with a thin sample of fused silica are presented and discussed. The discussion is first devoted to each parameter, and, afterwards, to their combination, which yields the final uncertainties of the indices. Finally, in Section 5, the conclusions are presented.

2. Foundations

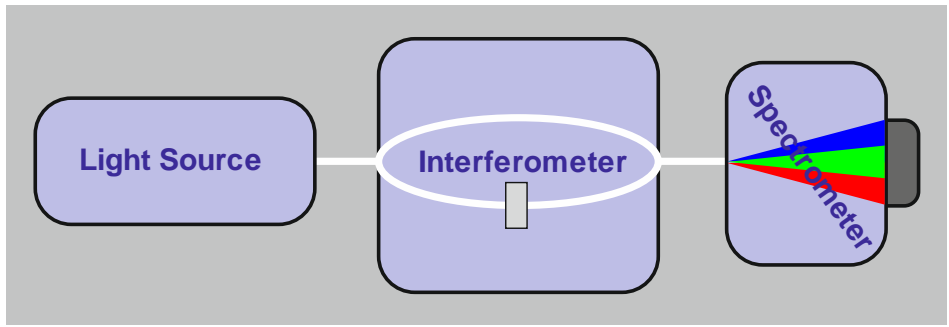

Fig. 1 Outline of a SRWLI device

In a SRWLI experiment (see Fig. 1), two broadband beams are superimposed, and the interference pattern is resolved in the spectral domain. At an arbitrary wavelength, $\lambda$, the irradiance can be written as:

$$I(\lambda) = I_0(\lambda)\left[1 + V(\lambda)\cos\varphi(\lambda)\right], \qquad (1)$$

with $I_0$ the background irradiance, $V$ the fringe visibility, and $\varphi$, the phase difference between the light beams. In the case in which one of the beams passes through a transparent plate of thickness $d$, the phase difference is given by:

$$\varphi(\lambda) = \frac{2\pi}{\lambda}\left[d(n - n_A) - n_A l\right], \tag{2}$$

where $n$ and $n_A$ are the plate and air refractive index, respectively, at wavelength $\lambda$, and $l$ is the path difference travelled by one of the beams with respect to the other, in air. In this expression, it is assumed that no path difference has been introduced other than that generated in the thin plate and in air propagation. It is important to stress that the computed phase differs from that given by the equation by a multiple of $2\pi$. This is because the arc cosine is a multivalued function or, in other words, different phases differing by a multiple of $2\pi$ result in the same value of the cosine. Therefore, when calculating the refractive index, we have an ambiguity that results from this lack of definition of the phase. Specifically, the refractive index at a wavelength $\lambda$ is calculated as

$$n = n_A + \frac{1}{d}\left(\frac{\lambda \varphi}{2\pi} + n_A l + k\lambda\right), \tag{3}$$

where $k$ is an unknown integer. Using this equation and the well-known formula for the propagation of uncertainties, we obtain the uncertainty of the refractive index as[1]:

$$\Delta n = \frac{1}{d}\sqrt{\left(\frac{\lambda}{2\pi}\Delta\varphi\right)^2 + (\lambda\Delta k)^2 + \left(\frac{n - n_{geq}}{\lambda}d\Delta\lambda\right)^2 + (n_A\Delta l)^2 + \left[(n - n_A)\Delta d\right]^2 + \left[(d+l)\Delta n_A\right]^2}. \tag{4}$$

In the third and fifth summands we have introduced $n$ to abbreviate the formula and $n_{geq}$ is the group index at the so-called equalization wavelength, that at which the irradiance has a stationary point. It fulfills:

$$l = d\left[\frac{n_{geq}}{n_{geq}\big|_{air}} - 1\right]. \tag{5}$$

As discussed in the introduction, in addition to refractive index, another parameter that can be calculated with RISBI is the group index:

$$n_g = \frac{d}{d\sigma}(n\sigma) = n + \sigma\frac{dn}{d\sigma}, \tag{6}$$

with $\sigma = 1/\lambda$, the wavenumber. From Eqs. (2) and (3), the following expression is obtained:

$$n_g = n_{gA} + \frac{1}{d}\left(n_{gA}l + \frac{1}{2\pi}\frac{d\varphi}{d\sigma}\right). \tag{7}$$

One way to compute the phase derivative is to determine the fringe periodicity. It is [29]:

---

[1] In the third term on the root, the following approximation was performed: $\left(n_g / n_{geq}\right)\big|_{air} \simeq 1$

$$\Lambda(\sigma) = \frac{2\pi}{|d\varphi/d\sigma|}, \tag{8}$$

and therefore:

$$n_g = n_{gA} + \frac{1}{d}\left(n_{gA}l \pm \frac{1}{\Lambda}\right), \tag{9}$$

where the modulus and the ± sign come from the fact that the phase derivative is zero at the equalization wavelength and changes its sign at each side: negative for shorter wavelengths, and positive for longer wavelengths.

The corresponding uncertainty is:

$$\Delta n_g = \frac{1}{d}\sqrt{(n_{gA}\Delta l)^2 + \left[(n_g - n_{gA})\Delta d\right]^2 + \left(\frac{1}{\Lambda^2}\Delta\Lambda\right)^2}. \tag{10}$$

Eqs. (4) and (10) are the key equations of this study. To determine the value of the refractive and group index uncertainty, each term will be analyzed separately. Note that in the case of double pass interferometers, as the Michelson interferometer, one of the beams crosses the sample twice, so in the previous equations $d = 2e$, where $e$ is the sample thickness.

3. Uncertainties

In this section, the uncertainties of the different parameters contributing to the refractive and group indices are discussed in detail. With respect to the uncertainty in the refractive index of air, an excellent review can be found in ref. [30].

3.1. Spectrometer calibration

The most common way to compute spectral irradiance is by means of a spectrometer. Here, no particular configuration will be considered, it is simply assumed that the resolution of the device is sufficient to resolve the interference fringes and that a linear camera is the spectrum sensing system. Calibration of the spectrometer consists in determining which wavelength falls in any pixel of the camera. In a typical calibration procedure, various sources of known wavelength illuminate the camera, to establish the correspondence between pixels and wavelengths. Then, interpolation, polynomial fitting, or physical modeling [31-36] are used to determine a relation between wavelength and pixel number. In this work, we do it by means of a WLSI-based approach [37].

If there is no sample in the interferometer, the spectral phase will be linear with the wavenumber $\sigma = 1/\lambda$. That means:

$$\varphi(N) = 2\pi l n_A \sigma(N) - 2k\pi, \tag{11}$$

where $N$ is the pixel number, and $l$ is, as before, the path difference in air travelled by the beams. In this calibration, a suitable path difference can be chosen to have many resolved spectral fringes. This equation can be inverted to express the wavenumber as:

$$n_A \sigma(N) = a\varphi(N) + b, \qquad (12)$$

with $a = 1/(2\pi l)$ and $b = k/l$. Now, the known values of the calibration wavelengths are used to perform a linear fit and determine the constants $a$ and $b$, providing the relation $\sigma(N)$ or $\lambda(N)$. Note that this fitting can be highly enhanced by choosing a path difference showing many resolved spectral fringes. Finally, Eq. (12) furnishes the uncertainty equation:

$$\Delta\sigma = \sqrt{(a\Delta\varphi)^2 + (\varphi\Delta a)^2 + \Delta^2 b}\Big/n_A, \qquad (13)$$

together with:

$$\Delta\lambda = \Delta\sigma/\sigma^2 = \lambda^2 \Delta\sigma. \qquad (14)$$

To know how to evaluate $\Delta\varphi$ in Eq. (13), see Section 3.4.

3.2 The path difference in air

The uncertainty of the path difference in air can be calculated by taking advantage of the linear relationship between phase and wavenumber in the absence of sample. Once the previous calibration was carried out, both magnitudes are related by the slope $a$ as:

$$l = 1/(2\pi a) \Rightarrow \Delta l = l\Delta a/|a|, \qquad (15)$$

where we apply the absolute value to compute the uncertainty. Although in this section we apply the same methodology as in the previous one, it is common for the path difference to be measured with a specific spectrometer of smaller spectral range than the one used to measure dispersion. This will be the case when the latter spectrometer cannot resolve the fringes associated with the measurement of $l$ (smaller than the sample thickness, but of the same order of magnitude).

3.3 The sample thickness

If the sample is solid, its thickness can be measured directly with a high-resolution gauge. We must take into account the instrument resolution, $\delta d_r$, the flatness, $\delta d_f$, and parallelism, $\delta d_p$, of the sample surfaces. The corresponding uncertainties are $\Delta d_r = \delta d_r / \sqrt{3}$, $\Delta d_f = \delta d_f / \sqrt{3}$, and $\Delta d_p = \delta d_p / \sqrt{3}$, respectively. In addition to these uncertainties, we must include the statistical uncertainty that results from carrying out the same measurement several times.

The uncertainty estimation of liquid samples is a little more complicated since they must be introduced into a transparent cell to be measured. You can proceed in two ways: (a)

Measuring the total thickness of the cell, the thickness of the walls and subtracting both magnitudes to obtain the internal thickness, *d*; (b) Directly measuring the internal thickness by applying RISBI to a liquid of known refractive index. In this case, the phase difference added by the walls must be subtracted. Typically, in both methods, the estimated uncertainty is twice that of solid samples.

3.4 Phase computation

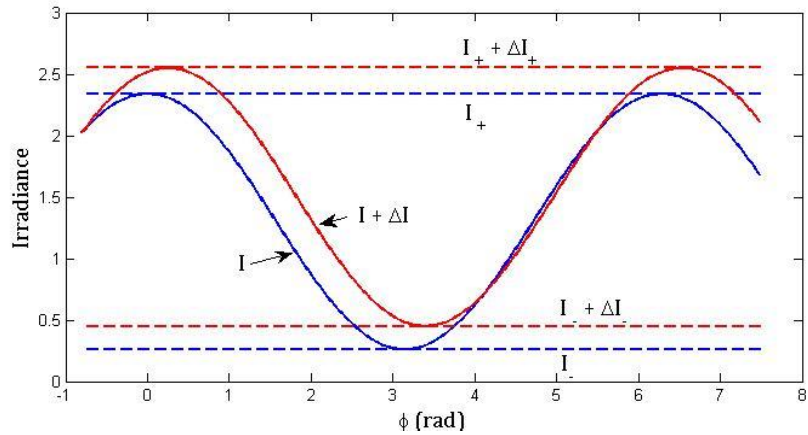

Fig. 2. Graph example of Irradiance variation

The calculus of the phase uncertainty is undoubtedly the most complicated step. It is determined by errors in the measurement of irradiance, but the latter also affects the background irradiance and visibility. To understand the effect of the irradiance variation on the phase, pay attention to Fig. (2). It plots the actual (*I*) and measured irradiance (*I+ΔI*) as a function of phase in one cycle and a bit. The oscillations are assumed to be very fast with respect to the background and visibility variation, so that we can keep the latter constant during a cycle. The irradiance variation affects both the upper and lower envelopes of the measurement as well as the phase. On the one hand, there is (or it can be) a change in the maximum and minimum values, which is associated to changes in the envelopes. On the other hand, there is also (or at least it could be) a lateral shift at any point of the interferogram, which is associated with a change in phase. Consequently, the uncertainty of the phase depends on the uncertainty of the irradiance and its envelopes. Each contribution can be determined separately.

Let us begin with the contribution of irradiance. The assumption that the envelopes are constant implies that the background irradiance and visibility are also constant. Therefore, from Eq. (1) it is obtained that the maximum phase variation due to irradiance is:

$$I + \Delta I = I_0 \left[ 1 + V \cos(\varphi + \Delta\varphi) \right], \tag{16}$$

or reusing Eq. (1):

$$\Delta I = I_0 V \left[ \cos\varphi (\cos\Delta\varphi - 1) - \sin\varphi \sin\Delta\varphi \right]. \tag{17}$$

In addition, we can approximate $\cos\Delta\varphi$ and $\sin\Delta\varphi$, so that to order $\Delta\varphi^2$ we have:

$$\Delta I = -I_0 V \left[ \cos\varphi (\Delta\varphi)^2 / 2 + \sin\varphi \Delta\varphi \right], \tag{18}$$

or, by reordering:

$$\cos\varphi (\Delta\varphi)^2 / 2 + \sin\varphi \Delta\varphi + \Delta I / (I_0 V) = 0. \tag{19}$$

The solutions to this equation are written as:

$$\Delta\varphi = \frac{-\sin\varphi \pm \sqrt{\sin^2\varphi - 2\cos\varphi \Delta I / (I_0 V)}}{\cos\varphi}, \tag{20}$$

where, for consistency, $\Delta I$ and $\cos\varphi$ have opposite sign and, therefore, the solutions are real. Although both solutions are correct, only certain branches of these remain bounded and are physically valid (see Fig. 3).

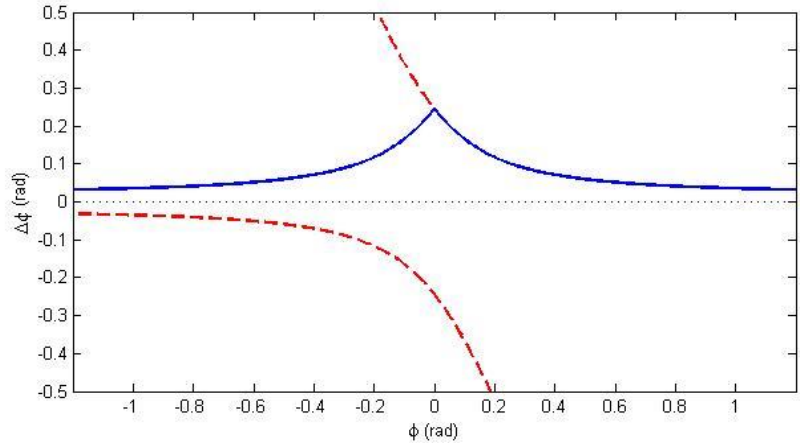

Fig. 3. Mathematical variation of the phase (dashed line) and the associated uncertainty (continuous line)

On the other hand, the uncertainty is taken as positive. Therefore, the physical solution for the uncertainty can be summarized in a single equation as:

$$\Delta\varphi_0 = |\tan\varphi| \left[ \sqrt{1 + 2|\cos\varphi| \Delta I / (I_0 V \sin^2\varphi)} - 1 \right], \tag{21}$$

where $\Delta I$ is assumed to be positive. This solution has two branches that meet at $\varphi = 0$ (see Fig.3), at which:

$$\Delta\varphi_0 = \sqrt{2\Delta I / (I_0 V)}, \tag{22}$$

value which corresponds to an extreme of the solution. On the other hand, at $\varphi = \pm\pi/2$, the solutions tend asymptotically to:

$$\Delta\varphi_0 = \Delta I / (I_0 V). \tag{23}$$

Under normal conditions $\Delta I/(I_0 V) < 1$, so that the maximum value of uncertainty corresponds to the one given in Eq. (22), while the minimum value is the one corresponding to Eq. (23).

In a second step, we must compute the contribution of the envelopes to the phase uncertainty. The envelopes, $I_\pm$, can be put as a function of background irradiance and visibility according to the following expressions:

$$I_\pm = I_0(1 \pm V), \tag{24}$$

and therefore, the irradiance can be recast as:

$$I = \frac{I_+}{2}(1+\cos\varphi) + \frac{I_-}{2}(1-\cos\varphi). \tag{25}$$

If we consider separately the variation of each envelope, we obtain:

$$I = \frac{I_\pm + \Delta I_\pm}{2}\left[1 \pm \cos(\varphi+\Delta\varphi)\right] + \frac{I_\mp}{2}\left[1 \mp \cos(\varphi+\Delta\varphi)\right]. \tag{26}$$

Proceeding similarly to the previous case, and after some simple calculations, we arrive at:

$$\cos\varphi(\Delta\varphi)^2/2 + \sin\varphi\Delta\varphi - \Delta I_\pm(1\pm\cos\varphi)/(2I_0 V) = 0, \tag{27}$$

which is similar to Eq. (19). Therefore, solving the equation we obtain the contribution of the envelopes to the phase uncertainty:

$$\Delta\varphi_\pm = |\tan\varphi|\left[\sqrt{1 + |\cos\varphi|(1\pm\cos\varphi)\Delta I_\pm/(I_0 V \sin^2\varphi)} - 1\right], \tag{28}$$

with $\Delta I_\pm$ positive. In particular, there is no contribution of the upper envelope to the phase uncertainty when $\varphi = (2m+1)\pi, m \in \mathbb{Z}$, and there is not contribution of the lower envelope when $\varphi = 2m\pi$. On the other hand, maximum uncertainty is $\Delta\varphi_\pm = \sqrt{2\Delta I_\pm/(I_0 V)}$. Putting all together and applying the law of propagation of uncertainties we obtain:

$$\Delta\varphi = \sqrt{\Delta^2\varphi_0 + \Delta^2\varphi_+ + \Delta^2\varphi_-}. \tag{29}$$

Examples of each relative contribution to the total phase uncertainty are shown in Fig. 4. As uncertainty decreases, the contribution of the irradiance approaches the sum of the contributions of the two envelopes.

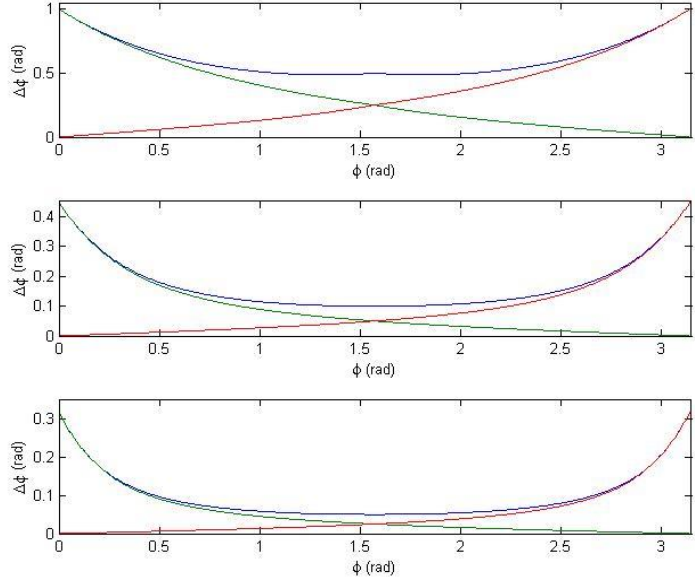

Fig. 4. Phase uncertainty related to irradiance $\Delta\varphi_0$ (blue), upper envelope $\Delta\varphi_+$ (green) and lower envelope $\Delta\varphi_-$ (blue). In each case, the irradiance uncertainty is constant and equal to $5 \times 10^{-1}$ (top), $1 \times 10^{-1}$ (middle) and $5 \times 10^{-2}$ (bottom).

3.5 Phase ambiguity

According to Eq. (3), the measured refractive index, here $n_k$, is a function of an arbitrary integer $k$ for a given wavelength and sample:

$$n_k = \frac{\lambda}{d}k + a, \tag{30}$$

with $a$ independent of $k$. Because of $k$ being an arbitrary integer, the raw refractive index measured with RISBI is also arbitrary in terms of $k$. The variation of the refractive index produced by two consecutive values of $k$ is:

$$\Delta n = \frac{\lambda}{d}. \tag{31}$$

That means that the error in the refractive index changes in steps of $\Delta n$. Therefore, in the case that the reference refractive index is known within a band smaller than $\Delta n$, the uncertainty associated to $k$ can be considered null. To overcome the ambiguity, the refractive index of the sample at a fixed wavelength, $\lambda_0$, is usually measured by a second method, which provides a way to determine $k$. In general terms, the result of measuring the reference refractive index by whatever procedure is $n_0 \pm \delta n$, where $n_0$ is the measured refractive index at the wavelength $\lambda_0$ and $\delta n$ its uncertainty. Hence, it can be said that the real value of the refractive index at $\lambda_0$ is contained in the interval $[n_0 - \delta n, n_0 + \delta n]$ with a high degree of confidence. This implies that we can get the correct value of $k$ provided that:

$$\delta n \leq \frac{\Delta n_0}{2} = \frac{\lambda_0}{2d}, \quad (32)$$

because in this case there is only one possible value of $n_k$ in the above interval $[n_0 - \delta n, n_0 + \delta n]$. In consequence, $n_k$ is the value that must be taken as correct instead of the measured reference refractive index $n_0$. On the other hand, if Eq. (31) is not fulfilled, the value of $k$ cannot be reliably assured. In view of this discussion, it can be said that the uncertainty in the value of $k$ is:

$$\Delta k = \text{floor}\left(\frac{2d}{\lambda_0}\delta n\right). \quad (33)$$

Here we assume that near $\lambda_0$, $\delta n \approx \delta n_0$. So, to reduce $\Delta k$ as much as possible, it is better to measure the reference refractive index at a longer wavelength. The $k$ uncertainty contributes to refractive index uncertainty as:

$$\frac{\Delta n}{\lambda} = \frac{1}{d}\text{floor}\left(\frac{2d}{\lambda_0}\delta n\right). \quad (34)$$

In Fig. 5 this contribution is plotted for $\lambda = \lambda_0$ as a function of the sample width, $d$.

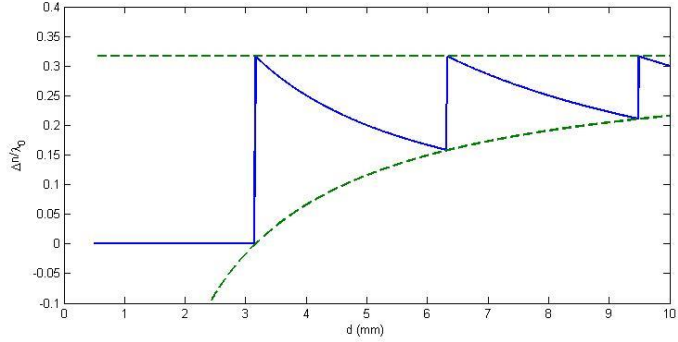

Fig. 5. Refractive index uncertainty normalized by the reference wavelength and caused by phase ambiguity against sample width. In this example, $\lambda_0 = 633$ nm and $\delta n = 1 \times 10^{-4}$.

For a fixed $\lambda$, $\Delta n$ makes a jump just when the argument of the floor function is an integer and, then decreases with $d$ until another jump arises. In this figure, the upper envelope is just $2\delta n/\lambda_0$, and the lower envelope is $2\delta n/\lambda_0 - 1/d$. So, independently of the value of $d$, the maximum contribution to the refractive index uncertainty at wavelength $\lambda$ is: $2\delta n\lambda/\lambda_0$.

3.6 Fringe spacing

The local fringe spacing, $\Lambda(\sigma)$, corresponds to the distance in wavenumber between two consecutive maxima or minima. It is also twice the distance between consecutive extremes (i.e., the absolute difference between a maximum and its adjacent minimum). So, to determine $\Lambda(\sigma)$, the extremes of the irradiance must be located. Calling $\sigma_\pm = \sigma \pm \Lambda/4$ the wavenumber for two consecutive minima and maxima, $\Lambda(\sigma)$ takes the form:

$$\Lambda(\sigma) = 2(\sigma_+ - \sigma_-) \quad \sigma = (\sigma_+ + \sigma_-)/2, \tag{34}$$

and for the uncertainty:

$$\Delta\Lambda(\sigma) = 2\sqrt{(\Delta\sigma_+)^2 + (\Delta\sigma_-)^2}. \tag{35}$$

By definition, irradiance maxima belong to the previously defined upper envelope, $I_+$, and irradiance minima belong to the lower envelope, $I_-$. As discussed above, a change in irradiance leads to a change in phase which, in turn, varies the wavenumber of the irradiance extrema. However, a change in the value of envelopes, does not affect the position of extrema but only affects to their value. So, the uncertainty in wavenumber can be written as (see Eq. 7):

$$\Delta\sigma = \frac{\Delta\varphi}{2\pi\left[dn_g - n_{gA}(d+l)\right]}, \tag{36}$$

where $\Delta\varphi$ is related to $\Delta I$ by Eq. (20). Note that once the period is determined using Eq. (34) for two extrema, it can be interpolated for any value of $\sigma$, and, for this reason, Eq. (36) applies to every $\sigma$. It can also be calculated directly from the irradiance pattern by noticing that:

$$\left[I(\sigma_+) - I_0(\sigma_+)\right]/V(\sigma_+) = \cos\left[\varphi(\sigma + \Lambda/4)\right] = -\cos\left[\varphi(\sigma - \Lambda/4)\right] = \left[I_0(\sigma_-) - I(\sigma_-)\right]/V(\sigma_-), \tag{37}$$

and taking care that $\sigma_\pm = \sigma \pm \Lambda/4$ when the uncertainty at $\sigma$ is computed using Eq. (35).

4. Results and discussion

In order to evaluate the impact of the different uncertainty sources on the retrieval of refractive and group indices over a broad spectral range, we analyze in this section measurements of standard experiments carried out in our laboratory. Although results are specific to our experimental setup, the analysis is general enough to be applicable to measurements with typical RISBI devices. The main features of our RISBI system have been described elsewhere [10]. It can perform real-time dispersion measurements in a broad spectral range, from 260 to 1650 nm. With the purpose of keeping the discussion as general as possible, the results shown in this work are restricted to the interval from 400 to 1000 nm, where only one single spectrometer is used. Despite RISBI can operate with just two spectral acquisitions, one of the interferogram and another of the background irradiance, we have repeated each one of them 10 times to have statistically meaningful measurements. Indeed, it could be possible to extract the background irradiance from the interferogram, but we prefer to acquire it separately, since it allows for a double-check of the validity of the data. In Fig. 6 we show how a standard measurement with our device looks like. As it is usual in our experiments, the interferogram contains the stationary phase point, which assures good visibility in the whole spectral range, even for high dispersive samples. In this section, we present results obtained with a fused silica optical window, 2 mm thick. Characteristics and manufacturer's specifications are detailed in Table 1.

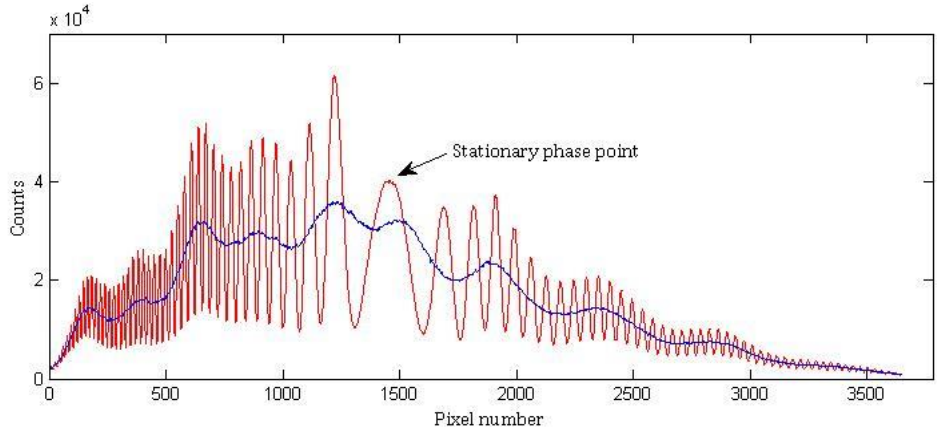

Fig. 6. Example of interferogram and background irradiance against pixel number

As in the preceding section, we analyze the impact on the measurement of each parameter separately.

Table 1. Sample parameters

| | |
|---|---|
| Diameter | 20.0 mm |
| Material | UV Grade Fused Silica |
| Antireflection Coating | Uncoated |
| Surface Flatness | λ/10 @ 632.8 nm |
| Surface Quality | 20-10 scratch-dig |
| Thickness | 2.0 mm |
| Thickness Tolerance | ± 0.1 mm |
| Wavefront Distortion | λ/10 @ 632.8 nm |
| Parallelism | <5 arc seconds |
| Clear Aperture | Central, 80% of diameter |
| Diameter Tolerance | +0.0/-0.2 mm |

## 4.1. Spectrometer calibration

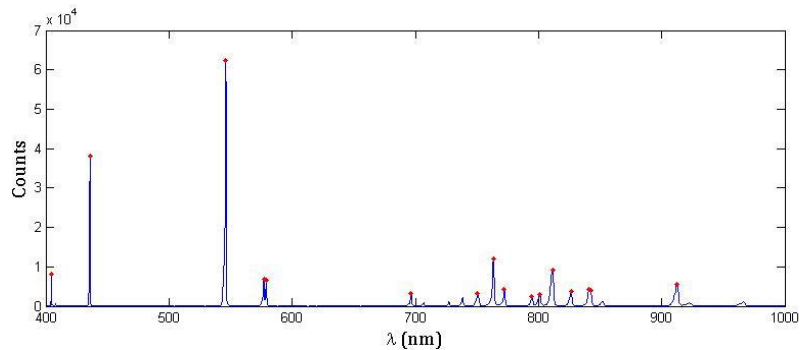

Fig. 7. Spectrum of a low-pressure Hg discharge lamp. Red points correspond to the lines used to perform the fit.

A low-pressure mercury discharge lamp was used to carry out the spectrometer spectral calibration, which is performed without sample in the interferometer. The spectra lines of this lamp in the range of measurement are shown in Fig. 7. Only the marked lines were used for calibration purposes. In order to remove any dispersion in the interferometer and ensure that induced phase difference is a linear function of the wavenumber in air, two interferograms corresponding to two different optical paths in air were taken and their phases subtracted. The computation of phase and its uncertainty was done using the method described in Section 3.4. Among the three sources of uncertainty (slope and intercept of the linear regression, and phase), the phase is the most important one. Hence, the calibration can be considered correct, but there is a need of improving the phase measurement (this will be discussed below, in Section 4.4). Fig. 8 shows that this phase uncertainty produces fast oscillations in the uncertainty of the wavelength and wavenumber within a period. In a real experiment, such a rapid change in uncertainty will be unrealistic and the phase is averaged over one cycle, which corresponds to the red line in Fig. 8(b).

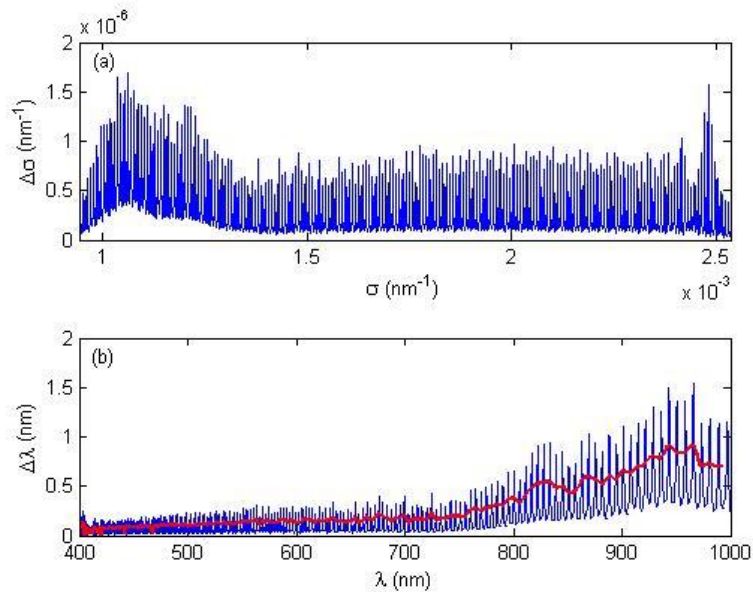

Fig. 8. Wavenumber and wavelength uncertainty. The red line is the local mean.

4.2 Path difference in air

The slope of the linear regression of the calibration process can be used to extract path difference $l$. The uncertainty of this magnitude is straightforward to obtain, and it is $\Delta l$ = 141 nm for a path difference of $l$ = 1.9860 mm. Notice that these magnitudes correspond to a Michelson interferometer where light passes through the sample twice and that it will be the half in a Mach-Zehnder type interferometer.

## 4.3 Sample thickness

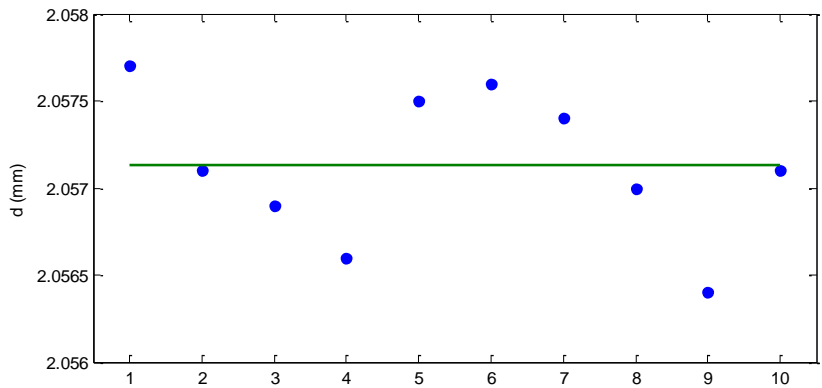

Fig. 9. Ten different measurements of the sample thickness, $d$, and their mean (green line).

The fused silica plate thickness was measured with a high accuracy micrometer with resolution of 100 nm. The measure was repeated 10 times to get its statistical mean and its standard deviation (see Fig. 9). To calculate the uncertainty due to the sample flatness and parallelism, a beam diameter of 5 mm and centered in the sample was considered. Results are shown in Table 2. Please, note that in the case of using a Michelson interferometer, light passes through the sample twice, so both, means and uncertainties of the table, have to be multiplied by a factor two.

Table 2. Sample thickness uncertainty and mean.

|  | Uncertainty (nm) | Mean (mm) |
|---|---|---|
| Resolution | 58 | - |
| Flatness | 9 | - |
| parallelism | 69 | - |
| Statistical | 135 | 2.05713 |
| Total | 160 | 2.05713 |

## 4.4 Phase

There are several sources of errors that contribute to phase uncertainty. On the one hand, mechanical and thermal vibrations vary the phase in the interferometer and so the interference pattern; on the other hand, lack of visibility and noise in the spectrometer deteriorate the image interferogram. In order to take into account the different contributions, we first took up to 10 interferograms and calculated their mean plus the standard deviation, as well as the mean and standard deviation of their envelopes (in blue, in Fig. 10). Furthermore, we applied a filter in the Fourier space and quantified the difference between the filtered and unfiltered interferograms (in green, in Fig. 10) as well as the difference of the corresponding envelopes. This gives an idea of the influence of noise. As a limit, absolute values of the spectrum smaller than one thousandth of the maximum value were filtered.

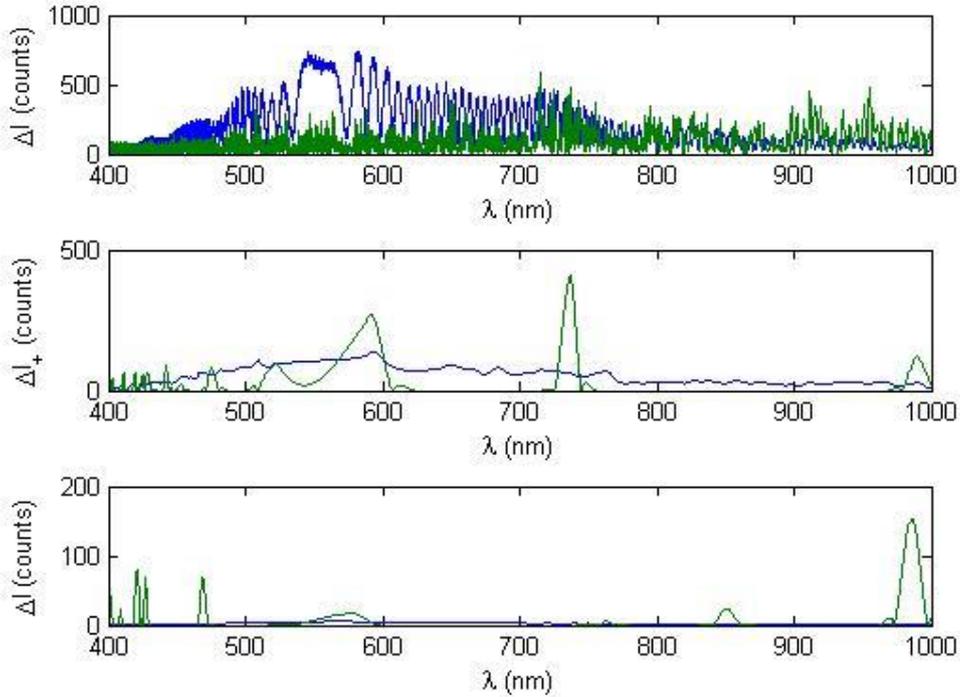

Fig. 10 Statistical (blue) and noise (green) contribution to irradiance (top), upper (medium) and lower (bottom) envelope uncertainty.

Both statistical and noise contributions to the uncertainty of the irradiance are noisy, however the statistical contribution is greater in magnitude. Regarding the uncertainty of the envelopes, it is mainly constant with large modulations. While we believe that those modulations at the border of the spectrum are caused by lack of visibility, we do not recognize any possible cause of those in the center of the spectrum but note that they are less than 1% of the envelope values. In Fig. 11 the three contributions to the phase uncertainty are separately shown. In agreement with the analysis in Section 3.4, all contributions present oscillations from period to period following Eqs. (21) and (28). The maxima and minima (zeros) of the phase contributions related to the upper and lower envelopes alternate at phase values multiple of $\pi$, coinciding with the maxima of the phase contribution related to the irradiance, while the minima of the later (greater than zero) occur when the phase is an odd multiple of $\pi/2$. Except at the spectrum borders, the greater contribution is related to the irradiance uncertainty, followed by the one related to the upper envelope.

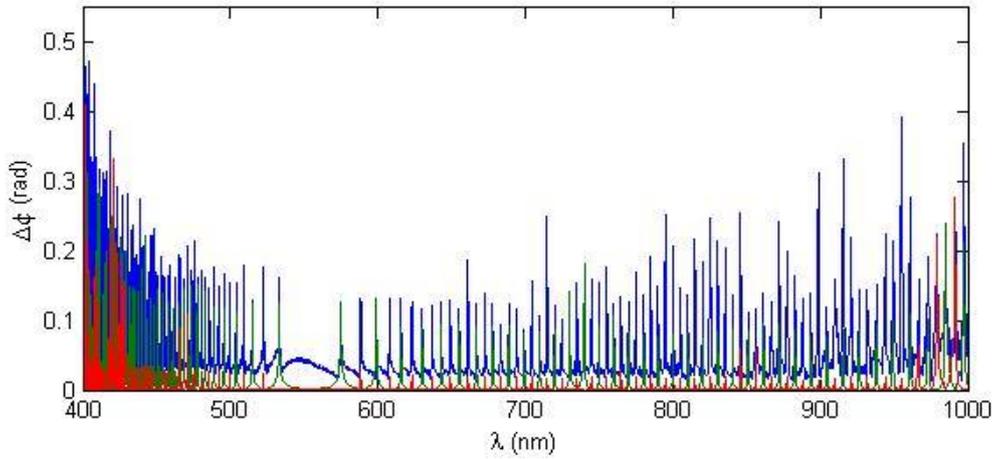

Fig. 11. Contributions to the phase uncertainty related to the irradiance (blue), upper envelope (green and lower envelope (red)

In Fig.12, which is taken as the final result of this section, the total phase uncertainty and its local average between 0.1 – 0.3 rad are shown.

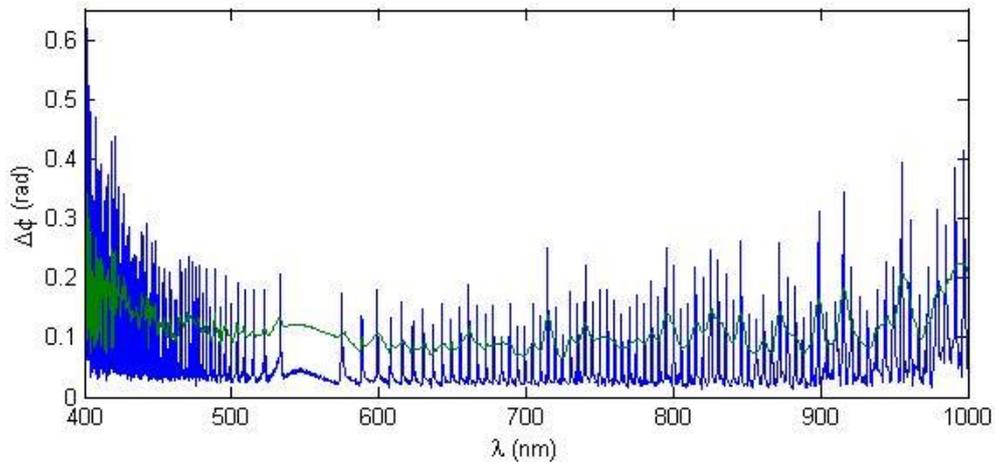

Fig. 12. Total phase uncertainty and its local average.

4.5 Phase ambiguity

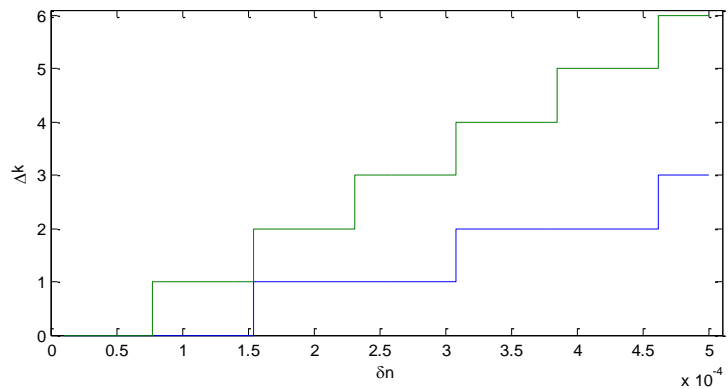

Fig. 13. Error in *k* as a function of uncertainty of the reference refractive index for a Mach-Zehnder (blue line) and Michelson (green line) type interferometer.

The uncertainty due to the phase ambiguity, $\Delta k$, depends on the sample thickness, the wavelength at which the reference refractive index is measured, $\lambda_0$, and the uncertainty of this measure, $\delta n$. In Fig. 13 we plot $\Delta k$ against $\delta n$ for our sample in both Michelson and Mach-Zehnder type interferometers, taking as reference the emission wavelength of a He-Ne laser, 632.8 nm. In the first case, there is no contribution of the phase ambiguity for $\delta n$ less than $7.7 \times 10^{-5}$ (the double, in the case of a Mach-Zehnder interferometer). This accuracy can be easily achieved by widely used refractometric techniques as those based on many interferometric approaches or minimum deviation methods. Indeed, commercial refractometers are available with an accuracy in the order of $5 \times 10^{-5}$ or better. We note that the impact of the phase ambiguity in the refractive index uncertainty increases with sample thickness, just the opposite to the contributions of phase, thickness, and path difference in air.

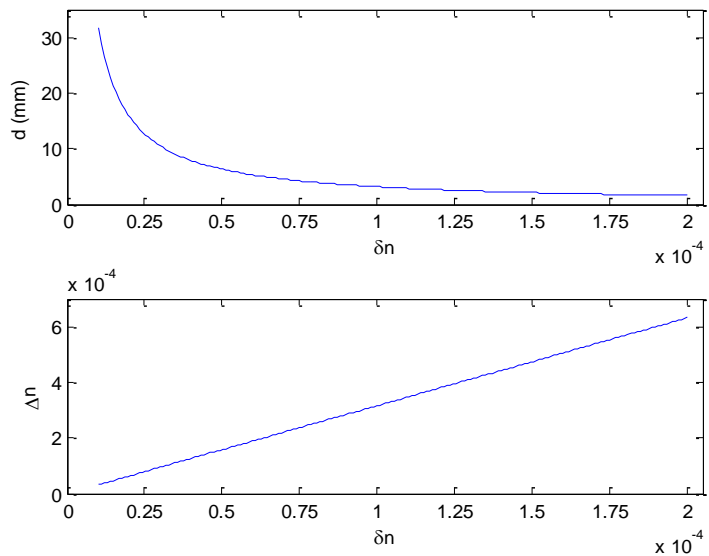

Fig 14. Largest *d* for which $\Delta k$ is cero as a function of $\delta n$ (top) and the contribution to refractive index uncertainty at a wavelength of 1 $\mu$m if $\Delta k$ = 1 (bottom).

In Fig. 14(a) it is shown the largest sample thickness for which the phase ambiguity contribution to the refractive index uncertainty is zero as a function of $\delta n$. In Fig. 14(b) it is plotted the contribution to the refractive index uncertainty of the minimum value of the sample thickness that produces $\Delta k$ = 1, at a wavelength of 1 μm, for a varying $\delta n$ too. As before, the reference wavelength is 632.8 nm.

4.6 Fringe spacing

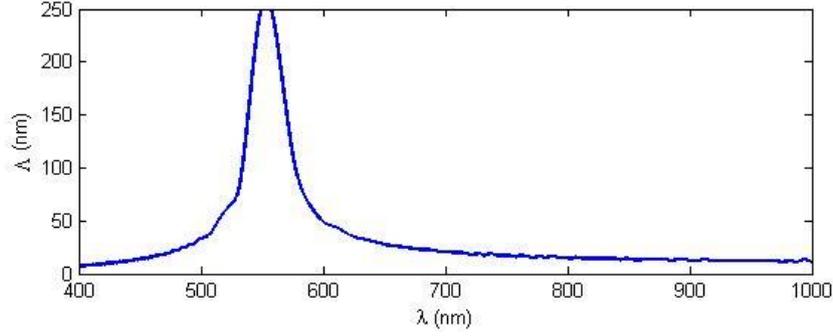

Fig. 15. Local periodicity as a function of wavelength.

As discussed in Section 3.6 the local periodicity is first calculated as $\Lambda(\sigma)=2(\sigma_+ - \sigma_-)$ with $\sigma_\pm = \sigma \pm \Lambda/4$, corresponding to maxima and minima of the interferogram. That means that $\sigma$ is close to a point where the irradiance equals the background ($\varphi = \pi/2$). Then, the periodicity is interpolated to cover all the measured points. The result is plotted in Fig. 15. The periodicity goes from approximately 10 nm at the spectrum borders to infinity at the equalization wavelength. Since the periodicity is proportional to the phase uncertainty, it shows oscillations as depicted in Fig. 16.

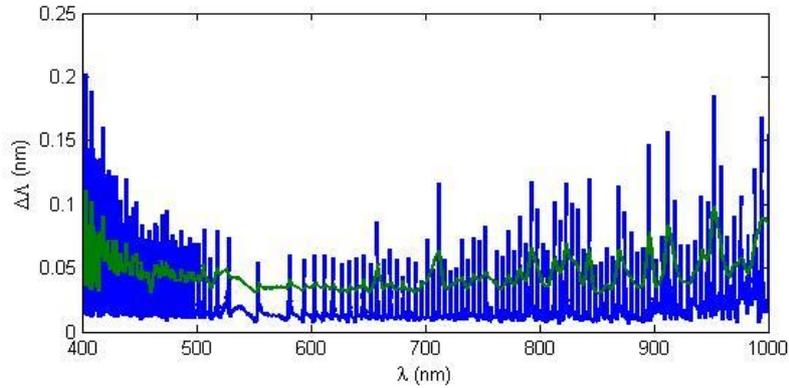

Fig. 16. Periodicity uncertainty and its local average.

4.7 Refractive and group index uncertainty

After individually analyzing the different sources of uncertainty to refractive and group indices, all these contributions are considered together in this section. Fig. 17 shows the total refractive index uncertainty as a function of the wavelength as well as the contribution of each of the uncertainty sources. We have assumed that the phase ambiguity is well resolved so its contribution to refractive index uncertainty is zero. Below in the text, some appointments were done about this issue. In addition, the uncertainty of air refractive index is not shown since, according to [38], it is orders of magnitudes smaller than the rest of contributions ($\approx 10^{-8}$).

Observing the graph, it can be concluded that the main contribution to the refractive index uncertainty results from the path difference in air, $l$, ($\approx 6.8 \times 10^{-5}$), followed by the sample thickness ($\approx 3.3 \times 10^{-5}$). The contribution of the calibration of the spectrometer is only relevant at IR wavelengths, whereas the phase contribution is negligible. That gives a total refractive index uncertainty nearly constant, in the range from $7.6 \times 10^{-5}$, at visible wavelengths, to $8.3 \times 10^{-5}$, at IR wavelengths.

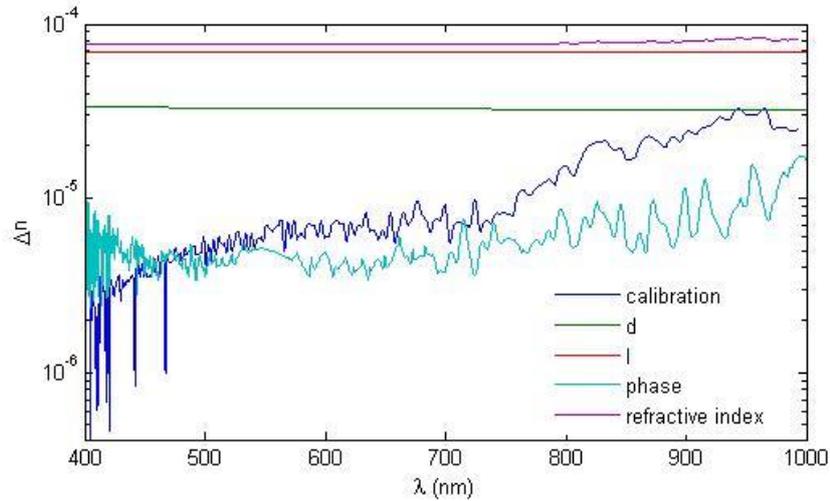

Fig. 17 The different contributions to the refractive index uncertainty and the resulting refractive index uncertainty

On the other hand, incorporating the contribution of the periodicity, the uncertainty of the group index can be computed. Fig. 18 shows the group index uncertainty together with the contributions of its different sources. In this case, the periodicity is the factor that most affects the group index uncertainty, except near the equalization wavelength, where it increases greatly. Regarding the contributions of path difference in air, $l$, and sample thickness, $d$, they are very similar to those to refractive index uncertainty, as can be appreciated comparing the expressions in Eqs. (4) and (10).

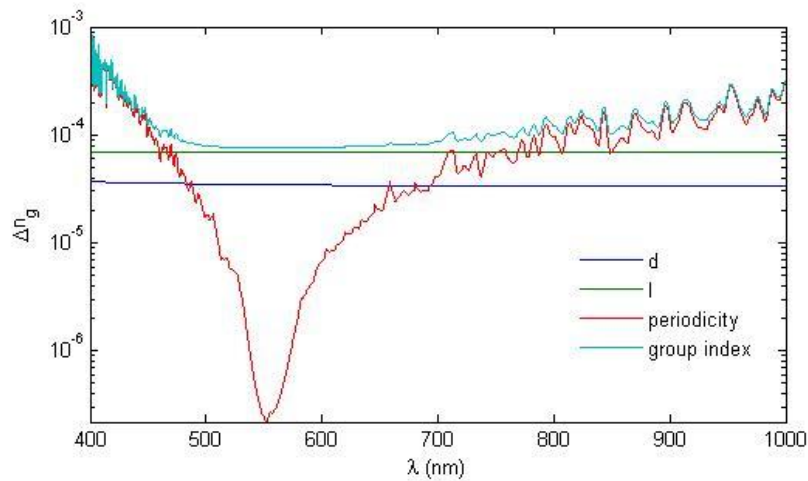

Fig. 18. The different contributions to the group index uncertainty and the resulting group index uncertainty

Finally, in Fig. 19, refractive and group index uncertainty are plotted together. The group index uncertainty exceeds the refractive index uncertainty in the whole spectrum, being both very similar near the equalization wavelength. At short wavelengths, the group index uncertainty reaches out values that are too high, of the order of $10^{-3}$.

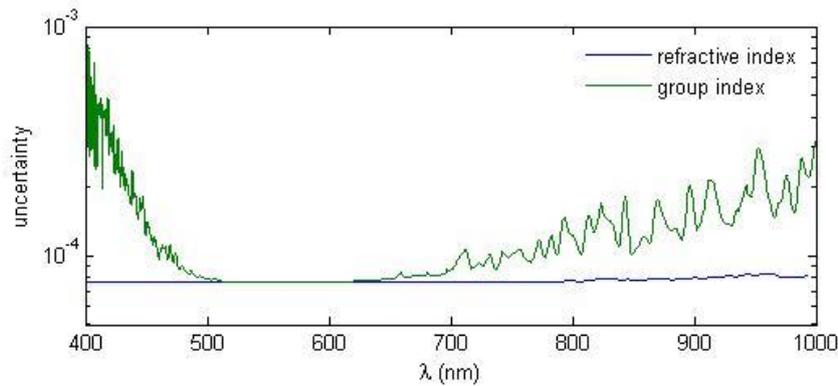

Fig. 19. Comparison between refractive and group index uncertainties.

4.7 The sample thickness again

We have seen that the two parameters that most effectively contribute to refractive index uncertainty are path difference in air, $l$, and sample thickness, $d$, both contributions decreasing with $d$. Therefore, we could think that by increasing $d$, the uncertainty of the refractive index could be further reduced. However, in that case, the contribution of the phase ambiguity ($k$) may not be negligible, and it may even exceed the previous ones if the reference refractive index is measured with not so good accuracy. In Fig. 20, we illustrate this effect by representing the total uncertainty of refractive index in addition to the contributions of $d$, $l$ and $k$ as a function of sample thickness. The considered wavelength is the Sodium D line, and the corresponding reference refractive index uncertainty is taken as $2 \times 10^{-4}$. In this case, the limiting value of the $k$ contribution is $4 \times 10^{-4}$, and it begins to be the main contribution for sample thickness greater than 1.47 mm (0.75 if the interferometer is a Michelson one). In addition, increasing $d$, also produces a reduction of the fringe periodicity of the interferogram which reduces visibility, and, collaterally, increases the phase uncertainty, which becomes the most relevant contribution to group index uncertainty.

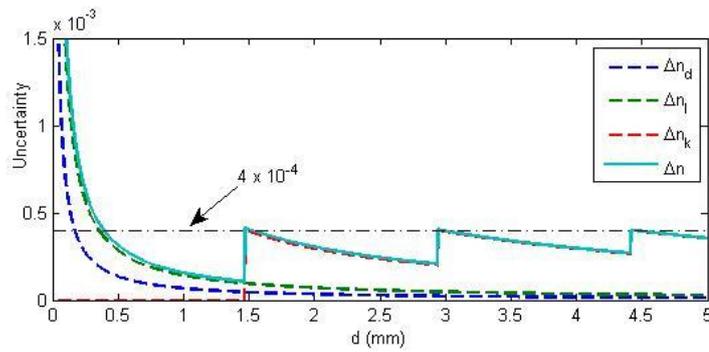

Fig. 20. Refractive index uncertainty against sample thickness. $\Delta n_d$, $\Delta n_l$, and $\Delta n_k$ correspond to $d$, $l$ and $k$ contribution, respectively.

5. Conclusions

It has been proved that difference of path in air, sample thickness measurements and phase ambiguity are the factors that mostly affect the accuracy of refractive index measures in RISBI. The uncertainty related to difference of path in air and sample thickness decreases with sample thickness. In turn, phase ambiguity cancels for very thin samples but as the thickness increases it can exceed any other contribution, being limited by the measure of the reference refractive index. In the example presented, a compromise between these different components gives a smaller uncertainty for sample thickness about 1 – 1.5 mm. With respect to the measure of group index, another parameter comes into play: the interferogram local periodicity or fringe spacing. Indeed, the measure of this parameters is the great cause of inaccuracy of group index measurement. While we have obtained refractive uncertainties less than $10^{-4}$, the attained group index accuracy must certainly be improved.

Acknowledgement

The financial support of the Spanish Ministry of Economy and Competitiveness through the coordinated grants MAT2017-89239-C2-1-P is gratefully acknowledged. Moreover, this work was funded by the Xunta de Galicia and FEDER (ED431D 2017/06, ED431E2018/08, GRC 508 ED431C 2020/10). Facilities provided by the Galician Supercomputing Centre (CESGA) are also acknowledged.